# E-learning model for art education in Iran

**Bahram Sadeghi Bigham**
Department of Computer Science and Information Technology, Institute for Advanced Studies in Basic Sciences (IASBS), Zanjan, Iran (b_sadeghi_b@iasbs.ac.ir)

**Mahboubeh Fannakhosrow**
Department of Curriculum, Farhangiyan University of Tehran, Tehran, Iran (mfannakhosrow@gmail.com)

**Aliakbar Safipour**
Education and Psychology Group, Department of Information Science and Knowledge , Alzahra university, Tehran, Iran (aliakbar.safipor@yahoo.com)

**Mostafa Jafari**
Department of Management and accounting, University of Zanjan, Zanjan, Iran (Jafari.mostafa@znu.ac.ir)

**Khadijeh Chenari**
Department of Mathematical Sciences, Alzahra university, Tehran, Iran (Kh.chenari@alzahra.ac.ir)

**Abstract**

Before the advent of Corona crises, virtual education was considered a viable alternative that was less popular in the arts for a number of reasons. But with the advent of the Covid virus 19, its use in business became necessary. According to the characteristics of art, art education according to Bloom's classification, should take place at high and close to the level of creativity. E-learning related to the subject of art is different from ordinary E-learning. In this paper, a general circular model for E-learning for art education is presented, a case study which has been done in Iran among students and artists. Art in the general sense, in addition to the same and Comprehensive features, has characteristics specific to different countries and cultures, and consequently there are differences in art education. This paper addresses the features of teaching art electronically (online or offline) in a specific culture and country (Iran) as the case study. The new model includes three main parts "Policymakers and planners", "Designers and manufacturers" and also "Supervisors, manufacturers and investors".

# 1-Introduction

Art is one of the most important and influential issues in human society, which often results in the production of works of art and, consequently, thought and culture. Art, like various sciences, is learned and combined with individual abilities and talents. Learning and teaching art has always been associated with concerns especially in Iran. The rich Iranian Islamic culture and art does not have a worthy educational background and this is the confession of the pioneers of art education. With the advent of digital tools, opportunities and sometimes emerging threats have emerged in the world of art education. In all countries and scientific-educational centers, and consequently in the subject of art education, measures were to be taken, which entered the world of e-learning in several stages, slowly and away from haste.

Entering education in the electronic world was welcomed in various sciences and environments. The reception in the sciences taught by lecturing was quite different from that in the world of mathematics and reasoning. But in the meantime, art was very different and it was much harder to work in the digital world for education. Art requires deep education (not to the extent of memorization), and on the other hand, art education is accompanied by practical work and exercises. Because of these features, the greats of science education sought to take advantage of the electronic world, but students and artists were less interested in it.

Facing the global crise of Corona left no choice for education and forced us into the digital world. For various reasons, the problems in the world of art education were more serious than other issues, four of which are discussed below.

1-1. **The need to learn at high levels with artistic subjects:** In Bloom's hierarchical pyramid of learning, there are six levels of learning; to know (to memorize), to understand (to describe and explain contents), to apply (practical use of what has been learned), to analyze (detailing the issues and revealing the motives and reasons), to compose (putting together the components and producing a new work) and to evaluate (judging the works). The essence of art is such that, according to Bloom's classification, education must be at least at the level of composition, which in itself is very deep.

1-2. **The difficulty of E-learning at higher levels of learning:** The more we enter the higher levels in Bloom's hierarchical pyramid, Education will become more difficult in the digital world. Learning in the first and second levels can happen easily by watching movies and lectures, but in the next levels, interactions between the teacher and the learner play a very important role, which in art is also one of the pillars of learning.

1-3. **More need for high quality sound, image, graphics:** In addition to the basic difficulties in E-learning in art, there are special requirements in this area. The quality of sound, images and any multimedia file is of particular importance, and reducing the quality of these exchanges will directly affect the quality of art education. In most educational subjects, these cases are not very important, and even disconnecting and connecting in sending low-quality video and images does not interfere with education.

**1-4. Insufficient mastery of students in digital educational tools:** Art teachers in Iran, in addition to confirming the above points, point out that students in recent years, have felt less need to use the digital world, and even with a strong educational platform is not ready to enter the world of digital education. This is also one of the important factors that make it difficult to enter the world of digital education with the subject of art.

Despite these problems, one cannot escape the world of digital education and its opportunities and threats. In a social demand, everyone does an important part of education digitally.

In this way, it countered the decline in the quality of works of art, the unemployment of students and schools, and its other destructive effects. The present plan is an attempt to structure E-learning (online and offline) and its ancillary requirements (E-exam, schools, teaching qualifications) in which a general model for E-learning is presented. The data and documents used as well as field studies are on the subject of art and this project is proposed for the Islamic Republic of Iran.

Building a country independently and on the basis of science is one of the goals that has been emphasized a lot. In today's world, countries must be able to live and interact in the digital world and be able to communicate with people and institutions in cyberspace. The subject of this project is the provision of very serious and important services in this new world that has always been emphasized and approved. But after the outbreak of Corona disease, this need is felt with a staggering leap, prompting officials to embrace and provide qualitative online education on a digital platform or information carriers. This is not just for government products and in the form of public services. Rather, senior education managers intend to prepare and present the necessary preparations and documents so that the engine of manufacturing digital products and providing services in the country by the private sector become more principled and with added value. These are all manifestations of the ability that is directly related to the economy and improving the livelihoods of people and artists. Before any new action and review, it is necessary to acquire knowledge about the quality and requirements of this field, and this knowledge must necessarily be valid and on par with global science.

On the subject of economics, much emphasis is placed on the country's independent economy, which is based on quality and mass production, equitable distribution, fair and extravagant consumption, and wise management relationships. The first two of these four cases are in line with the objectives of the present plan, and the most basic of these is the quality of products. If this plan is fully and correctly implemented, quality products of educational products will be presented on various platforms in the digital world, which will make art education in the society wider, cheaper and more qualitatively acceptable. Also, one of the next results of the project will be that the final productions of art in the country will grow significantly in terms of quantity and quality, and in this way we will be able to take steps in the issues of employment and entrepreneurship, non-oil exports and improving the living conditions of artists.

In a large part of the world of services, attention has been paid to various types of E-learning with the aim of teaching art in the country, which its results can be used for other scientific and educational topics on a digital platform. The plan will include a general model of E-learning based on three basic principles of education that pay special attention to art education topics. Attempts have also been made to provide a practical list that can be used to evaluate related products with the opinion of experts. The results of these evaluations can be used in future decisions (approval / disapproval, support, introduction, proposal for improvement, etc.) that enable those in charge, including the Office of Planning and Art Education, to make easier and fairer decisions.

The main issue is to respond more fundamentally to the needs of the art community, which for previous reasons (benefits of E-learning) and new requirements (Corona prevalence and the need to maintain job opportunities) should be considered a basic measure for teaching art courses in the digital world. These matters are currently being carried out on an island without coordination and at the same time with job inconveniences and poor haste, and it is necessary to review and coordinate in general by the government and with the support of the Office of Planning and Art Education.

At present, with the acceleration caused by Corona and due to the backwardness of some companies as well as the poor (but fast) production of some manufacturers, this is being done unfavorably and only to meet the essential needs. In some cases, illegal activities are carried out and some information of the country is misused. Regarding the importance of this type of study, it should be said that according to studies conducted in European countries, about 61% of E-learning courses are not suitable and lack the necessary efficiency. (Bari & Djoiab2014). It has also been shown that the failure rate of online courses is higher than face-to-face courses (Xu & Jaggers 2011). The reason for these events is issues similar to the situation in Iran during the Corona crises, which have inevitably entered E-learning without scientific principles. The same issues have been examined in other studies and in one of these documents it has been shown that the main reason for the inefficiency of E-courses is superficial training. (Kathleem, 2016). The project providers are wary of this type of training. The reason for these failures has been re-examined by others and briefly shown to be the inefficiency caused by the unplanned and direct transfer of classes from traditional to electronic without examining learning methods. (Thalheimer, 2017). Other studies have shown the need for an E-learning system that includes all elements of education. (Khan 2005). Of course, there are certain advantages to use training courses that make it attractive even without careful planning. These include adjusting the pace of learning for participants. (Adkins, 2014). But for most cases, such as participatory learning, a more accurate program must be presented in the electronic world. (BJekic et al. 2010). The proposal also includes other issues such as support, which is of particular importance (R. Nejhad et al. 2016, Smpson 2012)

However, the needs of society in the field of E-learning are increasing and this issue has become more important with the advent of Corona. The present project has been presented by examining various scientific aspects and experiences of other countries and also by receiving the opinion of some elites and studying scientific articles related to E- learning for art education in Iran.

Social conditions and legislation on the one hand and the hardware features and facilities and costs on the other hand are involved in the full implementation of this plan. But other important factors such as the three main pillars of education mentioned earlier and also teachers and art learners in Iran are the main reasons for the success or failure of the project.

## 2. E-learning

People use the term online learning in different ways. In general, we can refer to the method of providing educational information using the Internet and other tools in the digital world. This concept may include the presentation of digital textbook content and video or audio content through non-formal education or fully structured online courses with evaluation and certification. Online learning frees education from the constraints of time and space in face-to-face learning and can provide a more accessible form of learning for people looking for a wide range of educational opportunities. Although it is sometimes assumed that online learning and traditional classroom learning are at odds, this is not the case. In other words, learning on a digital platform is not copying the traditional world and repeating it in cyberspace. Online learning should be seen more as a different way of teaching and learning that can be used on its own or as a complement to classroom teaching. Likewise, online learning does not mean repetitive face-to-face teaching in an online environment. The power of online teaching and learning sometimes offers different and better learning experiences.

Formal online learning now uses the Internet. Therefore, educators need to be equipped with the Internet and desktop computers, laptops, tablets or other appropriate devices. It is often necessary to have an internet bandwidth in addition to this equipment.

E-learning and the use of cyberspace and online facilities provide a wide range of good facilities, but the lack of proper use of it sometimes causes a decrease in the quality and inability of education.The one-time use of the digital world for education, especially in times of crisis such as the Corona outbreak, has enabled any organization to conduct E-learning as quickly as possible, without paying attention to details and principles. It is not possible to move from traditional to E-learning without knowing the levels of learning, the differences in pedagogy and andragogy, the needs assessment techniques, the various teaching methods and the types of evaluation methods.The whole path of education (which is not limited to teaching) must be reopened to the new world. The present plan provides a model that can properly teach art in Iran and in the digital context to take advantage of digital tools and overcome its disadvantages or control it.This issue is much more important in the art world than other topics because art always requires creativity and creativity is one of the highest levels of Bloom's classification that can not be achieved simply by "remembering" and even "understanding" methods. The present study examines the similar models of some related countries as well as some similar training courses and proposes the final model for teaching different art courses in Iran through scientific methods.

The general model is proposed specifically for the Office of Planning and Art Education and includes many details to achieve general objectives. There are similar localized models in various documents that can only be cited as examples of the work of Iraklis and his colleague. (Iraklis & Loannis 2006). The depth of learning , attention to the cultural and artistic features , characteristics of the Islamic Republic of Iran , the emphasis on the compatibility of the system with the subject of art , and the defined existing standard in this field, are the features of this project.

## 2-1. A variety of E-learning practices

Here are two general modes of E-learning called synchronous and Asynchronous. Synchronous E-learning is the type of E-learning in which the parties interact simultaneously and learning occurs in the same way. This type of E-learning includes chat, video conferencing, software sharing and virtual classes. In the asynchronous type of E-learning, there is no need for the simultaneous presence of teacher and learner, and among these, we can mention different modes of self-learning, watching educational videos, learning in blog environments or learning while playing. But apart from this general category, E-learning can also be classified in other ways, which we will introduce in the following.

- **Self-study:** One of the most common learning methods in the digital world, where you can use, and learn existing documentation, which is often in pdf, doc, ppt or other similar formats.
- **Video and audio files:** The second most common method of E-learning that is used these days and can be presented in different ways. Sometimes they are presented on digital carriers and sometimes they are seen or heard online without the possibility of downloading.
- **Computer-Based Training (CBT) and Web-Based Training (WBT):** In this type of learning, instructional and educational content can be run and used on a compact disc or a file and program on a computer. In web-based mode, the same files and features can be used on the Internet, such as learning management systems or LMS. This type of educational learning is also a kind of self-reading method and there is no simultaneous interaction between the learner and the teacher.
- **Blended learning or instructor-led training (ILT):** is a combination of synchronous and asynchronous methods and is effective in cases where E-learning should have both synchronous and asynchronous aspects. In this type, training is often done asynchronously and by other methods, and online and simultaneous classes are used for exercises and interactions.
- **Learning with mobile or M-learning:** These days the most available communication tool is mobile, and education on this platform can be very useful and effective. For this type of learning, the design of training courses should be done in accordance with the mobile phone (specific screen size and memory and keyboard touch type, etc.).
- **Social learning:** is one of the most effective types of learning that learners learn from each other in an interactive environment. In these environments, users write their experiences and ask questions. The result of these interactions has great implications for learning that organizations often encourage their employees to participate in these types of job-related interactive environments.
- **Simulation:** One of the most practical and useful methods in E-learning, which reduces the cost of many experiments, trips and etc., and at the same time by using graphic tools and simulated laboratory environments, learning goes well.
- **Game-based learning or Gamification:** It is a clever type of training that encourages the learner to play and teaches him appropriate and sufficient content during the game. This type of education has become an academic discipline on its own, and many scholars in universities are currently working on this scientific subject.

In different sources, these categories are largely similar and sometimes slightly different. It has been seen that game-based learning is not considered as a separate type of learning and instead a method such as micro-learning is considered as one of the types of learning. In micro-learning, content is taught in completely separate and small packages (maximum ten minutes) and with a special structure.

## 2-2. Pillars of training and components of electronic model

Education thinkers believe that principled planning for education in any environment, age, subject and context requires a redefinition of principles in the three basic pillars. These three are needs assessment, education and evaluation, which in the present study and in the need assessment stage, the needs of the Iranian artistic community, including artists and students, as well as the social, economic and cultural needs of society are examined. At this stage, the requirements of the time and the culture and art of the country are considered, and in addition, the possibility of developing Iranian Islamic art and the discussion of exporting cultural and artistic works are considered. The next pillar is education, which is often seen as the basic pillar in which they deal with methods of teaching and content production. In this project, education is also of special importance and we try to consider the final model presented in all aspects and be optimal for art education in Iran and on a digital platform. It is emphasized that copying the traditional world and using it in the digital world has destructive effects on the education system, and according to the subject of art, teaching and learning should be done at the highest levels of Bloom's classification. In this section, the facilities available in the country as well as training organizations should be considered. The third pillar in this project is the study of different evaluation methods that can be easily implemented in the physical world with traditional methods. But when the tests are online, there are many problems to hold it fairly, which can not be solved with advanced features and high costs. For example, preventing fraud in drawing and uploading paintings is a case that can not be completely solved by spending a lot of money with webcams, internet, etc.

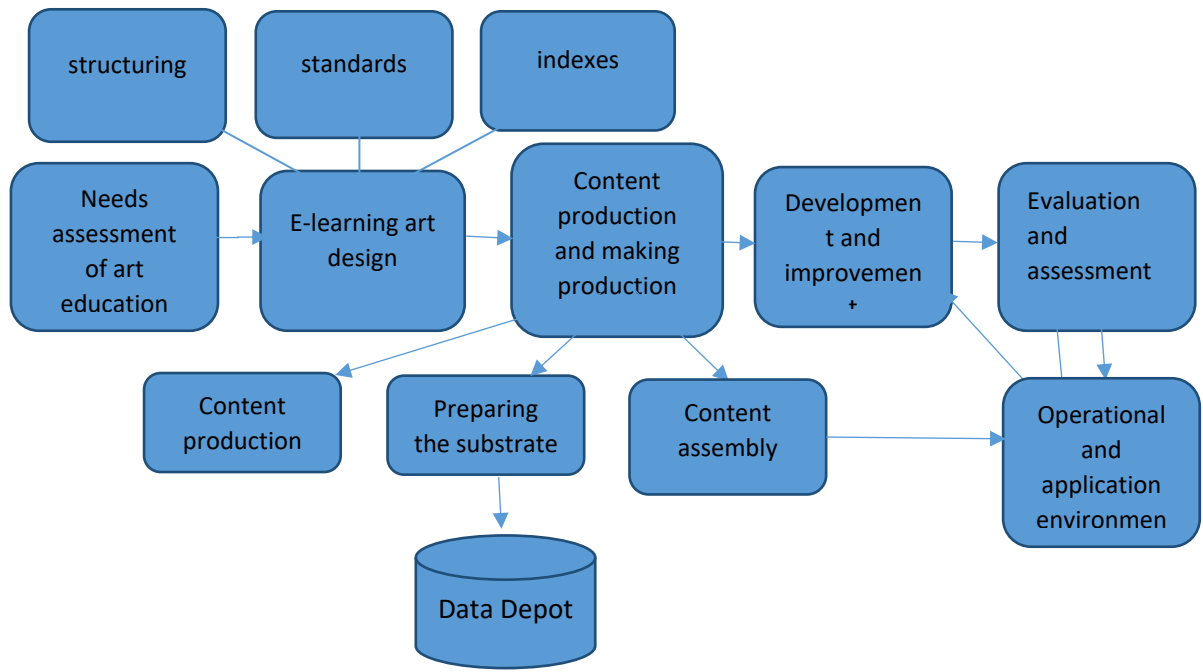

Figure 1: Outline of the components of the e-art education model

It is worth mentioning that the subject of student evaluation and evaluation of an educational system are fundamentally different but sometimes used interchangeably. In the present scheme, both terms evaluate an educational system of which student academic evaluation is a part. In the evaluation section, evaluation models of goal-based approach, management, consumer, expert opinion, expert disagreement, and naturalistic and participant-based approach have all been considered and examined in decision making and presentation of the final model.

**2-3. Outline of the basic components of the model**

According to the three basic pillars, the discussion of infrastructure, facilities and the process of related matters, the initial general plan is prepared in Figure 1, in which the general components and the process of affairs are specified. This general model has been done by pre-studying and receiving the opinions of experts in educational, artistic and E-learning affairs. The middle layer includes the main stages such as needs assessment, design and production, development and improvement, which require structuring, compliance with standards and indicators before production. Also, content production requires appropriate steps and platform, which is mentioned in the bottom layer along with the symbol of the repository. The third layer also shows the operating environment in which the product is being used and the feedback received from this environment can help develop and improve products. In the following, another general model will be presented as the main answer to the question of this plan, in which some of these steps are integrated and others are expressed in more detail. The final design of the pie chart shows three general steps with the relationship between them and each of the components (Figure 2).

The first step of any training model is training needs assessment. This means that we must first weigh and conclude what education society needs. How important is each training in terms of resource constraints and how much should they be valued? Extensive studies have been conducted in this regard and some results have been used as a global reference for many years. In the needs assessment section, economic issues and revenue generation and entrepreneurship goals, as well as national goals announced by policy makers, have a direct impact. Also, the cultural needs of the community and the needs expressed by artists and students can lead to correct recognition. The research results of Kaufman (1972) also provide a solution to the basic question of this section, and with the following category, we can also decide how to assess the needs of the project subject. According to the applications, the six types of educational needs assessment are as follows, which, after an overview, the method of assessment required in this plan is a combination of all cases.

- **Alpha Needs Assessment:** Considers reviewing, preparing, setting up, and implementing a variety of policies and guidelines. In alpha needs assessment, actions and activities are stated according to what they should be. It is not important to examine the possible disadvantages and problems. The main goal may even be to change current goals and objectives or to set new goals and objectives and develop new plans. This type of needs assessment introduces a definite path to identify changes and make them profound or starting point.

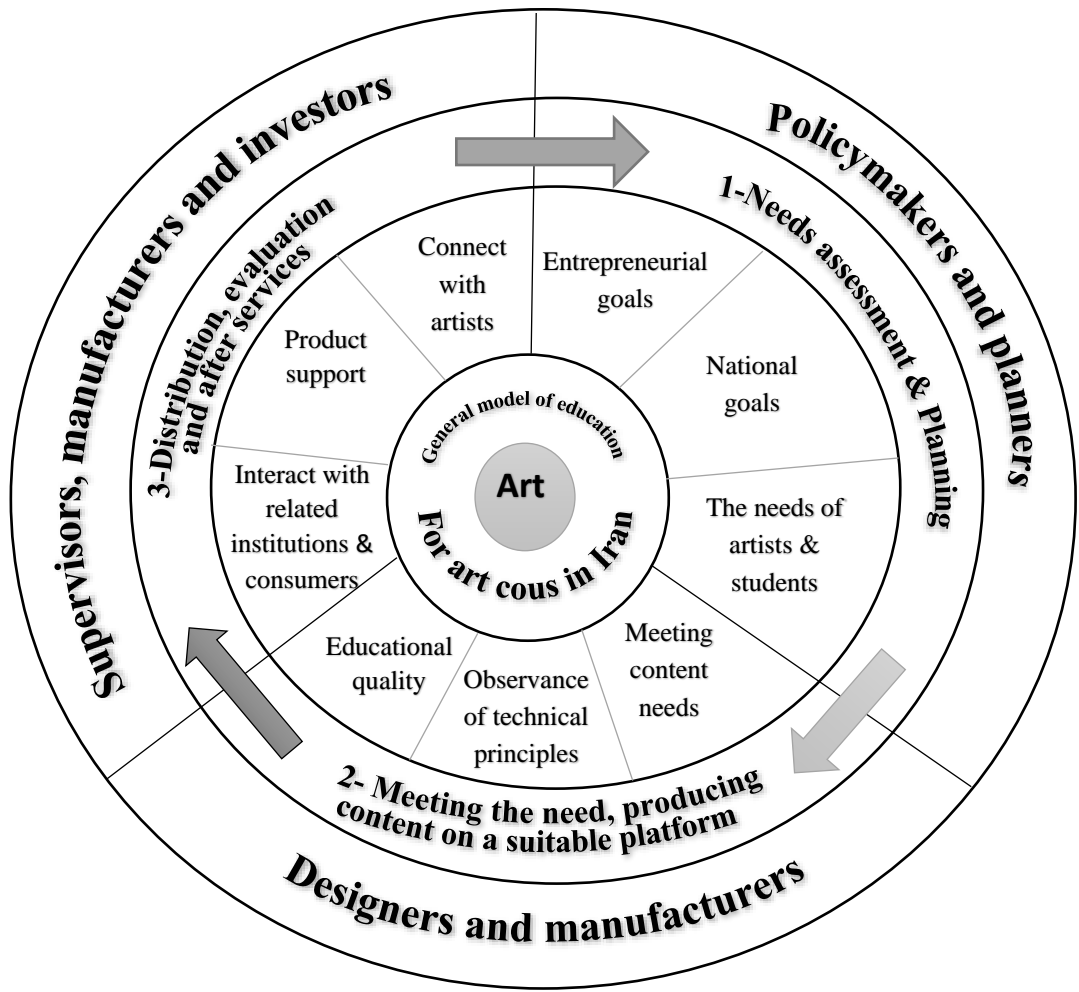

Figure 2: General model of e-learning for art courses

- **Beta Needs Assessment:** In this type of needs assessment, the goals, objectives, and training policies are assumed to be correct, and the purpose of the needs assessment is to identify distance or gap between current performance and desired performance.
- **Gamma Needs Assessment:** This type of needs assessment mainly focuses on prioritizing goals and objectives but their quality and quantity are not considered. In fact, prioritizing goals determines the amount of each need.
- **Delta Needs Assessment:** In this type of needs assessment, the various components of the actions are examined to determine the optimal performance of a task or duty to be determined. In other words, there is no discussion about the goals and general actions, but the details and the order of their actions and their relationship are examined and a needs assessment is done to identify the actions.
- **Epsilon Needs Assessment:** In this type of needs assessment, looking back, the difference between the results and the goals set is determined and analyzed. The results of this study will be used in the next steps and will be the basis for changing or modifying the program. Subsequent action needs are identified in this way and subsequent actions are taken accordingly.

- **Zeta Needs Assessment:** This type of needs assessment is a continuous review and always collects and analyzes information about the design, implementation and evaluation of the program. Based on these reviews and up-to-date data, new decisions are made about maintaining or modifying various components of the program. In this type of training needs assessment, a part of the program may change at some point without changing other components.

Considering the time position and previous actions in various fields of art education in Iran in both traditional and electronic methods, and taking into account all aspects and the emphasis in the second step statement, finally the alpha and beta method for needs assessment Art education is considered in the digital world. The stated basic objectives are examined and in addition, in production, the difference between it and the final objective is measured.

## 3. Representation in a suitable platform

Satisfying the identified needs is so important in the whole training process that sometimes other components of this system are ignored and production is seen as the whole process. In the matter of production, many points about the type of educational product (content, application, software, site, online education platform, ...) should be considered, each of which has its own various components. Each product must meet the standards announced in each country by policymakers or the private sector. This practice is different in different countries. In some countries, there are no fixed guidelines for it, and market tension and product attractiveness affect the game of market competition, forcing the manufacturer to follow certain principles. In some other countries, this is done by separate and sometimes parallel organizations.

In this part of the model, two separate and complementary perspectives prevail. One of them emphasizes the technical aspects of production and expects the manufacturer to present its product according to the correct technical standards. Accurate and flawless programming with appropriate execution time and acceptable storage volume, as well as ensuring its security and compliance of software needs with the tools available in the market and other similar items are some of the things that must be addressed in a significant part of production. At this stage, no attention is paid to the educational content and educational items and teaching methods and psychological impact, etc.

The second part emphasizes the production of educational content that is presented. Of course, if the product provides only a technical tool and lacks content, it can also be in the form of educational collections, but there will be no need to be examined in terms of educational content. But in most educational products that are presented with the subject of art, content has an important place. Therefore, it must follow the principles, the content must be correct, acceptable and can be measured by the criteria of virtual education.

These two basic categories in production need to be examined and quality assured. This means that a list of criteria must be prepared along with their degree of importance so that the product can be evaluated both during the production process and afterwards. Hence another subset of production in the overall model is considered as quality.

## 4. Conclusion and future work

In this paper, a general circular model for E-learning for art education presented. Also, we addressed the features of teaching art electronically (online or offline) in a specific culture and country (Iran) as the case study. The new model includes all the details in three main parts "Policymakers and planners", "Designers and manufacturers" and also "Supervisors, manufacturers and investors". For the future work, one can use the model to produce infrastructure and software (and also games) for e-learning purposes in art subjects. Also, the model can be used to evaluation system in every art education in the virtual world.